
\input harvmac.tex
\overfullrule=0pt

\def\simge{\mathrel{%
   \rlap{\raise 0.511ex \hbox{$>$}}{\lower 0.511ex \hbox{$\sim$}}}}
\def\simle{\mathrel{
   \rlap{\raise 0.511ex \hbox{$<$}}{\lower 0.511ex \hbox{$\sim$}}}}

\def\slashchar#1{\setbox0=\hbox{$#1$}           
   \dimen0=\wd0                                 
   \setbox1=\hbox{/} \dimen1=\wd1               
   \ifdim\dimen0>\dimen1                        
      \rlap{\hbox to \dimen0{\hfil/\hfil}}      
      #1                                        
   \else                                        
      \rlap{\hbox to \dimen1{\hfil$#1$\hfil}}   
      /                                         
   \fi}                                         %
\def\llgg{\ell^+ \ell^- \gamma \gamma}
\def\gggg{\gamma \gamma \gamma \gamma}

\def\jets{\rm jets}
\def\tpi{\pi_T}
\def\tpipm{\pi_T^\pm}
\def\tpip{\pi_T^+}
\def\tpim{\pi_T^-}
\def\tpiz{\pi_T^0}
\def\tpipr{\pi_T^{\prime 0}}
\def\toppip{\pi_t^+}
\def\mpi{M_{\pi_T}}

\def\pbarp{\ol p p}

\def\CM{{\cal M}}
\def\CO{{\cal O}}
\def\ts{\thinspace}
\def\ol{\bar}
\def\ra{\rightarrow}

\def\ev{{\rm eV}}
\def\kev{{\rm keV}}
\def\mev{{\rm MeV}}
\def\gev{{\rm GeV}}
\def\tev{{\rm TeV}}

\def\pb{{\rm pb}}

\def\fb{{\rm fb}}
\def\ecm{\sqrt{s}}

\def\half{\textstyle{ { 1\over { 2 } }}}

\def\twothirds{\textstyle{ { 2\over { 3 } }}}

\def\oneeight{\textstyle{{1\over {\sqrt{8}}}}}
\def\onetwelve{\textstyle{{1\over {\sqrt{12}}}}}
\def\onetwofour{\textstyle{{1\over {\sqrt{24}}}}}

\def\etmiss{\slashchar{E}_T}

\def\myfoot#1#2{{\baselineskip=14.4pt plus 0.3pt\footnote{#1}{#2}}}

\Title{\vbox{\baselineskip12pt\hbox{BUHEP-95-23}
\hbox{hep-ph/9507289}
}}
{Color--Singlet Technipions at the Tevatron}

\smallskip
\centerline{Kenneth Lane\myfoot{$^{\dag }$}{lane@buphyc.bu.edu}}
\smallskip\centerline{Department of Physics, Boston University}
\centerline{590 Commonwealth Avenue, Boston, MA 02215}
\smallskip

\vskip .3in

\centerline{\bf Abstract}

I discuss production and detection at the Tevatron $\pbarp$ collider of
pairs of light ($\mpi = 100$--$200\,\gev$) color--singlet technipions that
are expected in all nonminimal models of technicolor. Gluon fusion
production rates can be as large as $\CO(1\,\pb)$. Topcolor--assisted
technicolor is required to prevent top quarks from decaying as $t \ra \tpip
b$. An intriguing consequence of this is that the decays $\tpip \ra \tau^+
\nu_\tau$, $c \ol s$ and $\tpiz \ra b \ol b$ may also be suppressed so that
$\tpip \ra W^+ \gamma$ and $\tpiz \ra \gamma\gamma$ are significant. These
modes have spectactular signatures at the Tevatron.

\bigskip

\Date{7/95}

\vfil\eject

The focus of dynamical electroweak symmetry breaking has sharpened
lately. The emerging picture is based on topcolor--assisted
technicolor
\ref\tctwohill{C.~T.~Hill, Phys.~Lett.~{\bf 345B}, 483 (1995).},
\ref\tctwoklee{K.~Lane and E.~Eichten, {\it Natural Topcolor--Assisted
Technicolor}, BUHEP-95-11, FERMILAB-PUB-95/052, hep-ph/9503433 (1995),
to be published in Physics Letters~B.}.
In this picture, technicolor dynamics are largely responsible for
electroweak symmetry breaking
\ref\tcref{S.~Weinberg, Phys.~Rev.~{\bf D19}, 1277 (1979)\semi
L.~Susskind, Phys.~Rev.~{\bf D20}, 2619 (1979).}.
Extended technicolor interactions (ETC) still give rise to the hard
masses of quarks and leptons
\ref\etcrefs{S.~Dimopoulos and L.~Susskind, Nucl.~Phys.~{\bf B155}, 237
(1979)\semi
E.~Eichten and K.~Lane, Phys.~Lett.~{\bf 90B}, 125
(1980).}.
The unbroken technicolor interaction still must have a slowly running
coupling in order that fermion masses in the GeV range are produced while
large flavor--changing neutral current interactions are suppressed
\ref\wtc{B.~Holdom, Phys.~Rev.~{\bf D24}, 1441 (1981);
Phys.~Lett.~{\bf 150B}, 301 (1985)\semi
T.~Appelquist, D.~Karabali and L.~C.~R. Wijewardhana,
Phys.~Rev.~Lett.~{\bf 57}, 957 (1986);
T.~Appelquist and L.~C.~R.~Wijewardhana, Phys.~Rev.~{\bf D36}, 568
(1987)\semi
K.~Yamawaki, M.~Bando and K.~Matumoto, Phys.~Rev.~Lett.~{\bf 56}, 1335
(1986) \semi
T.~Akiba and T.~Yanagida, Phys.~Lett.~{\bf 169B}, 432 (1986).},
\ref\tasi{For a recent overview of technicolor, see K.~Lane, {\it An
Introduction to Technicolor},  Lectures given at the 1993
Theoretical Advanced Studies Institute, University of Colorado, Boulder,
published in ``The Building Blocks of Creation'', edited by S.~Raby and
T.~Walker, p.~381, World Scientific (1994).}.
However, it has been very difficult to generate a top quark mass
as large as that measured in the Tevatron collider experiments
\ref\toprefs{F.~Abe, et al., The CDF Collaboration, Phys.~Rev.~Lett.~{\bf
73}, 225 (1994); Phys.~Rev.~{\bf D50}, 2966 (1994); Phys.~Rev.~Lett.~{\bf
74}, 2626 (1995) \semi
S.~Abachi, et al., The D\O\ Collaboration, Phys.~Rev.~Lett.~{\bf
74}, 2632 (1995).},
even with ``strong'' extended technicolor
\ref\setc{T.~Appelquist, M.~B.~Einhorn, T.~Takeuchi and
L.~C.~R.~Wijewardhana, Phys.~Lett.~{\bf 220B}, 223 (1989);
T.~Takeuchi, Phys.~Rev.~{\bf D40}, 2697 (1989)\semi
V.A.~Miransky and K.~Yamawaki, Mod.~Phys.~Lett.~{\bf A4}, 129 (1989)\semi
K.~Matumoto, Prog.~Theor.~Phys.~{\bf 81}, 277 (1989) \semi
R.~S.~Chivukula, A.~G.~Cohen and K.~Lane, Nucl.~Phys.~{\bf B343}, 554
(1990).}.
Thus, ``topcolor'' interactions for the third generation quarks
seem to be required at an energy scale of about $1\,\tev$
\ref\topcondref{Y.~Nambu, in {\it New Theories in Physics}, Proceedings of
the XI International Symposium on Elementary Particle Physics, Kazimierz,
Poland, 1988, edited by Z.~Adjuk, S.~Pokorski and A.~Trautmann (World
Scientific, Singapore, 1989); Enrico Fermi Institute Report EFI~89-08
(unpublished)\semi
V.~A.~Miransky, M.~Tanabashi and K.~Yamawaki, Phys.~Lett.~{\bf
221B}, 177 (1989); Mod.~Phys.~Lett.~{\bf A4}, 1043 (1989)\semi
W.~A.~Bardeen, C.~T.~Hill and M.~Lindner, Phys.~Rev.~{\bf D41},
1647 (1990).},
\ref\topcref{C.~T. Hill, Phys.~Lett.~{\bf 266B}, 419 (1991) \semi
S.~P.~Martin, Phys.~Rev.~{\bf D45}, 4283 (1992);
{\it ibid}~{\bf D46}, 2197 (1992); Nucl.~Phys.~{\bf B398}, 359 (1993)\semi
M.~Lindner and D.~Ross, Nucl.~Phys.~{\bf  B370}, 30 (1992)\semi
R.~B\"{o}nisch, Phys.~Lett.~{\bf 268B}, 394 (1991)\semi
C.~T.~Hill, D.~Kennedy, T.~Onogi, H.~L.~Yu, Phys.~Rev.~{\bf D47}, 2940
(1993).}, \tctwohill.
As noted in Ref.~\tctwoklee, the main challenge facing topcolor--assisted
technicolor (TC2) is to provide a mechanism that generates large enough
mixing between the quarks of the third generation and those of the first
two.

The outline of this picture is sufficiently clear that it is worthwhile
to begin thinking about its phenomenology. In this Letter, I point out that
nonminimal models may have charged and neutral color--singlet
pseudo-Goldstone bosons---technipions, $\tpipm$ and $\tpiz$, light enough
to be pair--produced at the Tevatron $\pbarp$ collider and that, in TC2
models, these technipions may have unconventional decay modes that make
their detection easier. Until now, hadroproduction studies of these
technipions have concentrated on single production of weak--isosinglet
$\tpipr$ through gluon fusion and production of isovector pairs,
$\tpip\tpim$ and $\tpipm\tpiz$, via Drell--Yan processes
\ref\ehlq{E.~Eichten, I.~Hinchliffe, K.~Lane and C.~Quigg,
Rev.~Mod.~Phys.~{\bf 56}, 579 (1984); Phys.~Rev.~{\bf 34}, 1547 (1986).}.
Single--production of $\tpipr$ is invisible above backgrounds since the
decay modes $\tpipr \ra b \ol b$ and (independent of model) $gg$ are
dominant. The Drell--Yan rate for $\tpip\tpim$ is model--independent and
ranges from $30\,\fb$ down to $1\,\fb$ for $\mpi = 100$--$200\,\gev$.
However, if there are colored techniquarks, we shall see that
color--singlet technipions are pair--produced via gluon fusion with rates
that can be much larger than Drell--Yan. At the Tevatron, $\sigma(\pbarp
\ra \tpi\tpi) \simeq 1\,\fb$--$1\,\pb$, depending on details of the
underlying model. The upper end of this range is well within the Tevatron's
reach.

In standard technicolor models, $\tpip$ lighter than about $150\,\gev$
would be ruled out because the decay $t \ra \tpip b$ would dominate.
However, the $t \ra \tpip b$ coupling is proportional to the ETC--generated
portion of the top (and/or bottom) quark's mass and, in TC2 models, this is
at most a few~GeV. Thus, there is no danger of a large branching ratio for
this mode.\foot{Throughout this paper, I assume that there is no
significant mixing between $\tpip$ and the top--pion $\toppip$ associated
with spontaneous breaking of $t,b$ chiral symmetries.} Likewise,
constraints on technipions from the process $b \ra s \gamma$ are much
less stringent in TC2 models
\ref\balaji{B.~Balaji, {\it Technipion Contribution to $b \ra s \gamma$},
preprint BUHEP-95-18, hep-ph/9505313 (1995), submitted to Physical
Review~D.}.
Normally, fermionic final states would be expected to dominate $\tpi$ decays.
However, because third--generation fermion masses (including, possibly, the
$\tau$--lepton) are generated only in part by ETC interactions, this naive
expectation for the decays of a~$\tpi$ lighter than the top quark may be
wrong. An especially interesting possibility is that the anomalous
modes $\tpipm \ra W^\pm \gamma$ and $\tpiz \ra \gamma \gamma$ (but not $Z^0
\gamma$ for the isovector $\tpiz$) are significant
\ref\anomaly{J.~Ellis, M.~K.~Gaillard, D.~V.~Nanopoulos and P.~Sikivie,
Nucl.~Phys.~{\bf B182}, 529 (1981).}.
This can lead to striking signatures in $\pbarp$ collisions, especially
$\llgg + \etmiss$ and $\gggg$.

In the remainder of this Letter, I discuss
gluon--fusion of $\tpi\tpi$, estimate the production rates in two types of
model, and discuss the possibilities for $\tpi$ decays.
The scenarios considered are (1)~the one--family model
\ref\onefam{E.~Fahri and L.~Susskind, Phys.~Rev.~{\bf D20}, 3404 (1979).},
\ehlq\ and (2)~a multiscale model
\ref\multia{K.~Lane and E.~Eichten, Phys.~Lett.~{\bf 222B}, 274 (1989).},
\ref\multib{K.~Lane and M~V.~Ramana, Phys.~Rev.~{\bf D44}, 2678 (1991).}.
Bear in mind that these models have not yet been discussed in the context
of topcolor. In particular, the technifermions in these models do not yet
carry topcolor, as they must~\tctwohill,\tctwoklee. This will alter the
technifermion--content factors in the technipion states in Eqs.~(3) and~(6)
below and these, in turn, affect the production and anomalous decay rates.
Therefore, our estimates of these should be understood to be representative
of TC2 models.

There are two complementary mechanisms for gluon fusion of $\tpi$ pairs:
techniquark loops and colored technipion loops.\foot{This is implicit in
Ref.~\ref\bagger{J.~Bagger, S.~Dawson and G.~Valencia,
Phys.~Rev.~Lett.~{\bf 67}, 2256 (1991).}, where $gg \ra Z^0Z^0$ is
calculated. Also see Ref.~\ref\cgr{R.~S.~Chivukula, M.~Golden and
M.~V.~Ramana, Phys.~Rev.~Lett.~{\bf 68}, 2883 (1992).}.} The perturbative
techniquark loop calculation ought to be reliable when the typical momentum
flowing through the loop is large compared to the techniquark mass (and
large enough to be in the asymptotically--free regime of the walking
technicolor model). However, this regime is far above the $\tpi\tpi$
threshold region where the largest production rates are expected.

The other extreme is the chiral limit in which the $gg \ra \tpi\tpi$
amplitude is dominated by loops involving massless colored technipions.
Indeed, these massless particles produce a singularity that promotes the
amplitude from one involving gradiently--coupled external $\tpi\tpi$ to one
of effectively scalar--coupled $\tpi\tpi$
\ref\chiral{J.~F.~Donoghue, B.~R.~Holstein and Y.~C.~Lin, Phys.~Rev.~{\bf
D37}, 2423 (1988) \semi
A.~G.~Cohen, G.~Ecker and A.~Pich, Phys.~Lett.~{\bf 304B}, 347 (1993).}.
As the mass of the internal technipions is increased, they start to
decouple---as do heavy internal techniquarks. Consequently, the technipion
loop calculation should provide a reasonable approximation to the amplitude
in the threshold region, even in the case of broken chiral symmetry.

The amplitude for gluon fusion of longitudinal $Z$~bosons, $gg \ra Z^0_L
Z^0_L$, with massive intermediate color--triplet and octet technipions, has
been calculated in Ref.~\ref\eetl{E.~Eichten and T.~Lee, FERMILAB-PUB, in
preparation, (1995). I thank E.~Eichten for a preliminary version of this
paper.}. Modifying the group--theoretic factors there for the case of
interest to us, the amplitude for both $\tpip\tpim$ and $\tpiz\tpiz$ final
states is given by
\eqn\ggpp{\eqalign{
&\CM(g_a(q_1)g_b(q_2) \ra \tpi(p_1)\tpi(p_2)) =
{\alpha_S \over {8\pi F^2_T}} \ts \epsilon^\mu(q_1) \epsilon^\nu(q_2) \ts
\left({q_{2\mu} q_{1\nu} - q_1\cdot q_2 g_{\mu\nu} \over {q_1\cdot
q_2}}\right) \delta_{ab} \cr
&\ts\ts\ts \times \left\{ \sum_{R=3,8} T(R) \left[C_R \left(s -
\twothirds(M^2_R + \mpi^2)\right) + D_R \right]
\left(1 + 2 I(M^2_R,s) \right) \right\} \ts. \cr}}
Here, $F_T$ is the technipion decay constant; $s = (q_1 + q_2)^2$; $T(R) =
\half$ for $R= {\bf 3}$ and 3~for $R = {\bf 8}$; the factors $C_R$ and
$D_R$ are listed in Table~1 for the one--family and multiscale models; and
\eqn\zint{\eqalign{
I(M^2,s) &\equiv  \int_0^1 dx dy \ts {M^2 \over {xys - M^2 + i\epsilon}} \ts
\theta(1-x-y) \cr
&=\cases{-M^2 /2s \left[ \pi - 2 \arctan \sqrt{4 M^2/s -1}
\right]^2 & for $s <  4M^2$ \cr
M^2/2s \left[ \ln \left({1 + \sqrt{1 - 4 M^2/s} \over
{1 - \sqrt{1 - 4 M^2/s}}}\right) - i\pi\right]^2  &for $s > 4M^2$ \ts.} \cr}}

The one--family technicolor model~\onefam\ contains one doublet each of
color--triplet techniquarks $Q_{L,R} = (U,D)_{L,R}$ and color--singlet
technileptons $L_{L,R} = (N,E)_{L,R}$. These have the same electroweak
charge assignments as a quark--lepton generation.\foot{I assume throughout
that isospin breaking in the technifermion sector may be neglected.} All
technifermions transform according to the same complex irreducible
representation of the technicolor gauge group $SU(N_{TC})$. To ensure that
the technicolor gauge coupling ``walks'', this representation may need to
be larger than the fundamental.

The technipions of interest to us in the one--family technicolor model are
(the color index $\alpha=1,2,3$ is summed over):
\eqn\tpisinglets{\eqalign{
&|W^+_L\rangle =\half |U_\alpha \ol D_\alpha + N \ol E \rangle \cr
&|Z^0_L\rangle =\oneeight |U_\alpha \ol U_\alpha - D_\alpha \ol D_\alpha +
N \ol N - E \ol E \rangle \cr
&|\tpip \rangle =\onetwelve |U_\alpha \ol D_\alpha -3 N \ol E \rangle \cr
&|\tpiz\rangle =\onetwofour |U_\alpha \ol U_\alpha - D_\alpha \ol D_\alpha
- 3(N \ol N - E \ol E)
\rangle \ts . \cr}}
The decay constant of these technipions is $F_T = 120\,\gev$ (assuming
that for the topcolor pions $F_t = 50\,\gev$; see
\tctwohill). The $\tpip$ and $\tpiz$ masses arise mainly
from ETC interactions. In the one--family model, the ETC masses of
technifermions are comparable to those of the lighter quarks and leptons.
Thus, following the $\mpi$ calculations in Refs.~\multia\ and \multib,
\eqn\tpimass{
\mpi^2 = {2 m_T \over {F^2_T}} \langle \ol T T \rangle \simeq
8 \pi m_T F_T \simle (125\,\gev)^2}
for a technifermion hard mass $m_T \simle 5\,\gev$. Technipion masses in
the range 100--200$\,\gev$ are considered in this paper. Note that
color--triplet and octet technipions receive a further contribution to
their mass from QCD interactions and will be heavier. Typical octet and
triplet states are ($a=1,\dots,8$)
\eqn\tpicolor{\eqalign{
& |\tpi^{a,+} \rangle = \sqrt{\half}(\lambda_a)_{\alpha\beta}\ts
 |U_\alpha \ol D_\beta\rangle \ts, \cr
& |\tpi^{a,0} \rangle = \half(\lambda_a)_{\alpha\beta}\ts
|U_\alpha \ol U_\beta - D_\alpha \ol D_\beta\rangle \ts , \cr
& |\pi^\alpha_{U \ol E}\rangle = |U_\alpha \ol E\rangle \ts. \cr}}

Multiscale technicolor was invented~\multia\ as a way to implement a
walking technicolor gauge coupling~\wtc. The model considered here is a
simplified version of the one studied in Ref.~\multib. One doublet of
color--singlet technifermions, $\psi$, belonging to a higher dimensional
representation of $SU(N_{TC})$ is responsible for most of electroweak
symmetry breaking. The decay constant of the $\ol \psi \psi$ technipions
typically is $F_\psi = 220$--$235\,\gev$ (allowing for $F_t \simeq
50\,\gev$). I assume here that the light--scale technifermions consist of
one doublet each of techniquarks~$Q$ and technileptons~$L$, transforming
according to the fundamental representation of $SU(N_{TC})$. These condense
at a much lower energy with typical technipion decay constants $F_T \equiv
F_Q \simeq F_L = 30$--$50\,\gev$. The color--singlet isovector technipions
in this model are$^1$
\eqn\multipi{
|\pi_{Ti}^{\pm,0}\rangle = \sum_{F = \psi,Q,L} \gamma_{i F} | \pi_{\ol F F
}^{\pm,0} \rangle \qquad (i = W, Q, L)\ts .}
Here, e.g., $|\pi_{TW}^{\pm,0}\rangle = |W_L^{\pm,0}\rangle$ and the mixing
factors for these states are $\gamma_{W\psi} = F_\psi/F_\pi \simge 0.9$,
$\gamma_{WQ} = \sqrt{3}F_Q/F_\pi$, and $\gamma_{WL} = F_L/F_\pi$, where
$F_\pi = 246\,\gev$. As discussed in \multib, the diagonal mixing factors
$\gamma_{QQ} \simeq \gamma_{LL} \simge 0.9$ and off-diagonal ones are
small. In other words, the technipions are nearly ideally--mixed. This will
simplify the discussion below of their production and decays. The colored
technipions are still given by Eq.~\tpicolor. In the multiscale model, $Q$
and $L$ get their ETC mass from the heavy $\psi$ while $q,\ell$ get theirs
from $Q,L$. Then, the current masses of $Q,L$ will be much larger than
$5\,\gev$ and, despite the smaller $\ol Q Q$ and $\ol L L$ condensates,
$\mpi$ can easily be in the range 100--$200\,\gev$.

The total cross sections for $\pbarp \ra \tpip\tpim$ in the one--family and
multiscale models (using the factors~$C_R$ and~$D_R$ in Table~1) are shown
in Fig.~1. For the multiscale model, $F_Q = F_L = 40\,\gev$ was used. The
colored technipion masses were taken to be $M_3 = 200\,\gev$ and $M_8 =
250\,\gev$. The EHLQ Set~1 parton distribution functions~\ehlq\ were used
and the lowest order cross sections multiplied by~1.5 as an estimate of
radiative corrections to the gluon fusion process. For equal mass $\tpiz$
and $\tpip$, production rates for $\tpiz\tpiz$ are half as large as those
shown in Fig.~1.

At the Tevatron, the production rates for the technipions of the
one--family model are only a few femtobarns and it is unlikely that they
will ever be observable there, even with high--luminosity upgrades. I
believe that this sets the lower bound on $\tpi\tpi$ production by gluon
fusion. On the other hand, rates for the multiscale technipions with the
parameters used here range from 0.2 to $0.7\,\pb$. Depending on $\tpi$
decay modes, these could be large enough to observe with data from the
current run.\foot{The integrated luminosity accumulated by the CDF and D\O\
detectors at the end of Tevatron collider run~1 should be near
$125\,\pb^{-1}$.} It is important to remember that production rates vary as
$F_T^{-4}$ and increase fairly rapidly with decreasing $M_3$ and $M_8$. For
example, lowering $M_3$ to $175\,\gev$ and $M_8$ to $200\,\gev$ increases
the rates by a factor of 2--3, depending on $\mpi$.

An interesting feature of $gg \ra \tpi\tpi$ is shown in Fig.~2 where the
invariant mass distribution $d\sigma/d\CM$ is plotted for $\mpi =
110\,\gev$ and the cases $M_3 = M_8 = 0$ and $M_3 = 200$, $M_8 =250\,\gev$.
The distribution in the case of massless colored technipions is typical of
most pair--production processes in hadron colliders: it peaks a few 10s of
GeV above threshold and then falls off rapidly. In the massive case, the
distribution has sharp maxima at $\CM = 2 M_3$ and $2 M_8$, corresponding
to on--shell production of colored $\tpi\tpi$. Thus, for any reasonable
choice of masses, the color--singlet technipion invariant mass distribution
peaks considerably farther above threshold than would be expected for most
QCD processes. Observing this effect is tantamount to discovering the
colored technipions.

Let us turn now to the $\tpi$ decay modes. Technipions are expected to couple
to quarks and leptons with stength $m_{q,\ell}(\mpi)/F_T$, where
$m_{q,\ell}(\mpi)$ is the ETC--generated part of the fermion's mass,
renormalized at $\mpi$. Assuming that mixing factors suppress
inter--generational decay amplitudes, the dominant $\tpi$~decay rates are
expected to be\foot{I used $m_c(100\,\gev) = 0.3\,\gev$ and
$m_b(100\,\gev) = 2.5\,\gev$.}
\eqn\tpifbf{\eqalign{
& \Gamma(\tpip \ra c \ol s) \simeq {3 \over{16\pi}} {m_c^2(\mpi) \over
{F_T^2}} \mpi \simeq 0.3\kappa_T\,\mev \ts, \cr
& \Gamma(\tpip \ra \tau^+ \ol \nu_\tau) \simeq {1 \over{16\pi}}
{m_\tau^2(\mpi) \over
{F_T^2}} \mpi = 4\kappa_T\,\mev \ts, \cr
& \Gamma(\tpiz \ra b \ol b) \simeq {3 \over{16\pi}} {m_b^2(\mpi) \over
{F_T^2}} \mpi \simeq 25\kappa_T\,\mev  \ts, \cr}}
where $\kappa_T =(40\,\gev/ F_T)^2 \ts (\mpi/ 100\,\gev)$.

The hadronic decay channels in $\tpi\tpi$ production would be difficult to
detect. Without heavy flavor tagging, they are impossible to see above the
four--jet background. At least two heavy quarks must be tagged to bring the
$\tpiz\tpiz$ signal close to the $b b \ts + \ge 2 \ts \jets$ background and
this will entail a heavy acceptance loss. The $\tau + {\rm dijet} +
\etmiss$ signature has a large background from $W (\ra \tau \nu) + {\rm
dijet}$ production. The $\tau^+ \tau^- + \etmiss$ signal is impossible to
reconstruct and is swamped, e.g., by ordinary Drell--Yan ($Z^0 \ra
\tau^+\tau^-$) production. Fortunately, this unpromising scenario is not
the only possibility.

Technipions have another decay mode, namely, to two electroweak gauge
bosons. The amplitude for this decay is given by the triangle anomaly and
the rates are~\anomaly
\eqn\tpibb{\Gamma(\tpi \ra B_1 B_2) = (1 + \delta_{B_1 B_2}) \ts {S^2(\tpi B_1
B_2) \ts p_B^3\over {512 \pi^5 F_T^2}}\ts. }
The anomaly factor is
\eqn\sfact{S(\tpi B_1 B_2) = \half g_1 g_2 {\rm Tr}\left(Q_{\pi_T}
\left\{Q_{B_1}, Q_{B_2} \right\}\right) \ts,}
where $g_i$ is the $B_i$ gauge coupling and $Q_{B_i}$ is its (vector or
axial--vector) charge. For the isovector $\tpi$, the anomalous decay modes
are $\tpipm \ra W^\pm \gamma$, $\tpiz \ra Z^0 \gamma$ and $\tpiz \ra \gamma
\gamma$, with the factors given in the one--family model by
\eqn\tpisfact{\eqalign{
&S(\tpi W \gamma) = {N_{TC} \ts e^2 \over{\sqrt{6} \sin\theta_W}}
\ts,\cr
&S(\tpi Z \gamma) = {N_{TC} \ts e^2 \over{\sqrt{6}
\sin\theta_W \cos\theta_W}} (1 - 4 \sin^2\theta_W) \ts,\cr
&S(\tpi \gamma \gamma) = {4 N_{TC} \ts e^2 \over{\sqrt{6}}} \ts.\cr}}
The factors for the multiscale model are related to these
by $S(\pi_{\ol Q Q} B \gamma) = - {1\over \sqrt{3}} S(\pi_{\ol L
L} B \gamma) = \half S(\tpi B \gamma)$. These formulas assume that
techniquarks and leptons belong to the fundamental representation of
$SU(N_{TC})$. As noted earlier, technifermions of the one--family model
may need to belong to a higher representation to ensure a walking
technicolor coupling. This would enhance their anomalous decay rates.

It is clear why these decay modes generally have been considered
unimportant.\foot{Two counterexamples are: Randall and Simmons, who
made up for the small anomalous rates by considering $\tpi\tpi$ production
at the LHC~\ref\randsimm{L.~Randall and E.~H.~Simmons, Nucl.~Phys.~{\bf
B380}, 3 (1992).}; Lubicz and Santorelli, who considered multiscale $\tpiz$
production in $e^+e^-$ annihilation and assumed an {\it ad hoc} suppression
of the fermionic decay modes~\ref\lubicz{V.~Lubicz and P.~Santorelli, {\it
Production of Neutral Pseudo-Goldstone Bosons at LEP-II and NLC in
Multiscale Walking Technicolor Models}, BUHEP-9-16, hep-ph/9505336
(1995).}.} For $N_{TC} = 4$, $F_T = 40\,\gev$ and $\mpi = 100\,\gev$, the
anomalous rates are $3\,\kev$ for $W\gamma$, $2\,\ev$ for $Z^0\gamma$ and
$400\,\kev$ for $\gamma\gamma$, much smaller than the rates in Eq.~\tpifbf.
However, there is reason to be cautious about this conclusion: Eq.~\tpifbf\
may overestimate $\Gamma(\tpip \ra \tau^+ \ol \nu_\tau)$  and $\Gamma(\tpiz
\ra b \ol b)$. The fermion masses in these formulas are the ETC--generated
parts only. In TC2, the $b$--quark mass is due in part to ETC interactions
and topcolor instantons~\tctwohill. This $b$--mass, as well as the ETC mass
of the $\tau$, is then magnified to some extent by the topcolor and strong
$U(1)$ interactions
\ref\magnify{C.~T.~Hill and D.~Kominis, private communications. Also,
D.~Kominis, {\it Flavor--Changing Neutral Current Constraints in
Topcolor--Assisted Technicolor}, BUHEP-95-20, hep-ph/9506305 (1995).}.
Thus, the ETC masses of both the $b$ and $\tau$, like that of the $t$, may
be much less than their pole masses. Whether this is sufficient to suppress
the technipions' fermionic decay rates to the level of their anomalous
ones---ETC masses less than about $50\,\mev$ are required---and yet not run
afoul of phenomenological $b$~and $\tau$~constraints are model--dependent
issues that cannot be settled in this paper.

Spectacular signatures are associated with the anomalous decays of
technipions. For $\tpip\tpim \ra W^+W^- \gamma\gamma$, the cleanest signals
are $\ell^+\ell^-\gamma\gamma +\etmiss$, where $\ell = e$ or $\mu$ and all
invariant mass combinations tend to be large. These modes occur 1/81 of the
time for $ee\gamma\gamma$ and $\mu\mu\gamma\gamma$ and twice this often for
$e\mu\gamma\gamma$. For $\sigma \sim 1\,\pb$, it is hard to think of
backgrounds to these signals. Unfortunately, these clean modes cannot be
unambiguously reconstructed. For that, one must resort (as in top--pair
production, Ref.~\toprefs) to events in which one~$W$ decays hadronically,
giving the signal $\ell^\pm \gamma\gamma \ts + 2 \ts \jets + \etmiss$.
These occur 24/81 of the time. However, there will be losses due to
isolation and other cuts. It is important that the experimentalists
carefully simulate their acceptance for these events. The $\tpiz\tpiz \ra
\gggg$ mode is clear and unambiguous, though also subject to losses. In any
case, it is possible now to set meaningful limits on the charged and
neutral color--singlet $\tpi\tpi$ cross section-times-branching ratio with
existing data from the Tevatron collider. We hope this will become part of
the current search for new phenomena at the Tevatron.

\bigskip

This research was inspired by conversations with Estia Eichten. I thank
Sekhar Chivukula, Andy Cohen, Chris Hill and Dimitris Kominis for valuable
discussions and the Fermilab Theory Group for its hospitality during the
early stages of this work. This research is supported in part by the
Department of Energy under Grant~No.~DE--FG02--91ER40676.

\vfil\eject

\listrefs

\centerline{\vbox{\offinterlineskip
\hrule\hrule
\halign{&\vrule#&
  \strut\quad#\hfil\quad\cr\cr
height4pt&\omit&&\omit&&\omit&&\omit&&\omit&\cr\cr
&\hfill Model \hfill&&\hfill $C_3$ \hfill&&\hfill
$C_8$  \hfill&&\hfill $D_3$\hfill&&\hfill $D_8$\hfill &\cr\cr
height4pt&\omit&&\omit&&\omit&&\omit&&\omit&\cr\cr
\noalign{\hrule\hrule}
height4pt&\omit&&\omit&&\omit&&\omit&&\omit&\cr\cr
&1--Family$\tpip\tpim$&&\hfill${10\over{3}}$\hfill&&\hfill${1\over{3}}$\hfill&&
\hfill${16\over {9}} M_3^2$\hfill&&\hfill ${4\over{9}} M_8^2$\hfill&\cr\cr
\noalign{\hrule}
height4pt&\omit&&\omit&&\omit&&\omit&&\omit&\cr\cr
&Multiscale $\pi^+_{\ol Q Q} \pi^-_{\ol Q Q}$
&&\hfill${8\over{3}}$\hfill&&\hfill${4\over{3}}$\hfill&&
\hfill${32\over {9}} M_3^2$\hfill&&\hfill ${16\over{9}} M_8^2$\hfill&\cr\cr
\noalign{\hrule}
height4pt&\omit&&\omit&&\omit&&\omit&&\omit&\cr\cr
&Multiscale $\pi^+_{\ol L L} \pi^-_{\ol L L}$
&&\hfill$8$\hfill&&\hfill$0$\hfill&&
\hfill${16\over {3}}(2\mpi^2- M_3^2)$\hfill&&\hfill $0$\hfill&\cr\cr
height4pt&\omit&&\omit&&\omit&&\omit&&\omit&\cr\cr}
\hrule\hrule}}
\medskip

\medskip

\noindent Table 1. The factors $C_R$ and $D_R$ in Eq.~\ggpp\ for $gg \ra
\tpi\tpi$ for the one--family and multiscale technicolor models.

\vskip1.5truein

\centerline{\bf Figure Captions}
\bigskip

\item{[1]} The $\pbarp \ra \tpip\tpim$ production rates at $\ecm = 1.8\,\tev$
as a function of $\mpi$. The curves are $\pi^+_{\ol Q Q} \pi^-_{\ol Q
Q}$ (solid) and $\pi^+_{\ol L L} \pi^-_{\ol L L}$ (dashed) for the
multiscale model; $\tpip\tpim$ times 100 (dashed--dotted) for the one--family
model.

\item{[2]} The  $\pi^+_{\ol Q Q} \pi^-_{\ol Q Q}$ invariant mass
distribution for $\mpi = 100\,\gev$ in $\pbarp$ collisions at $\ecm =
1.8\,\tev$. The solid curve is determined with intermediate colored
technipion masses $M_3 = 200\,\gev$ and $M_8 = 250\,\gev$; the dashed curve
uses $M_3 = M_8 = 0$.

\vfil\eject

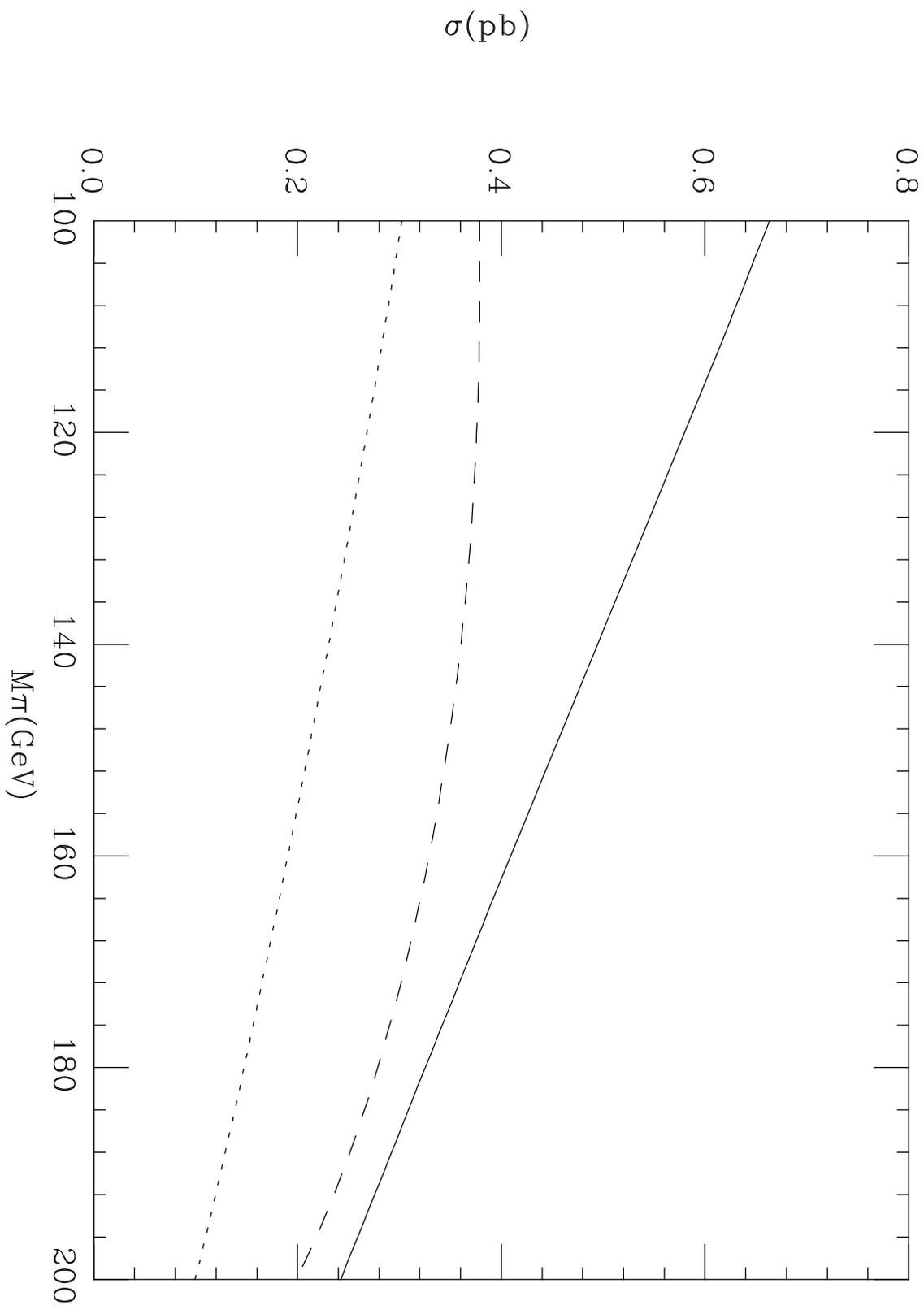
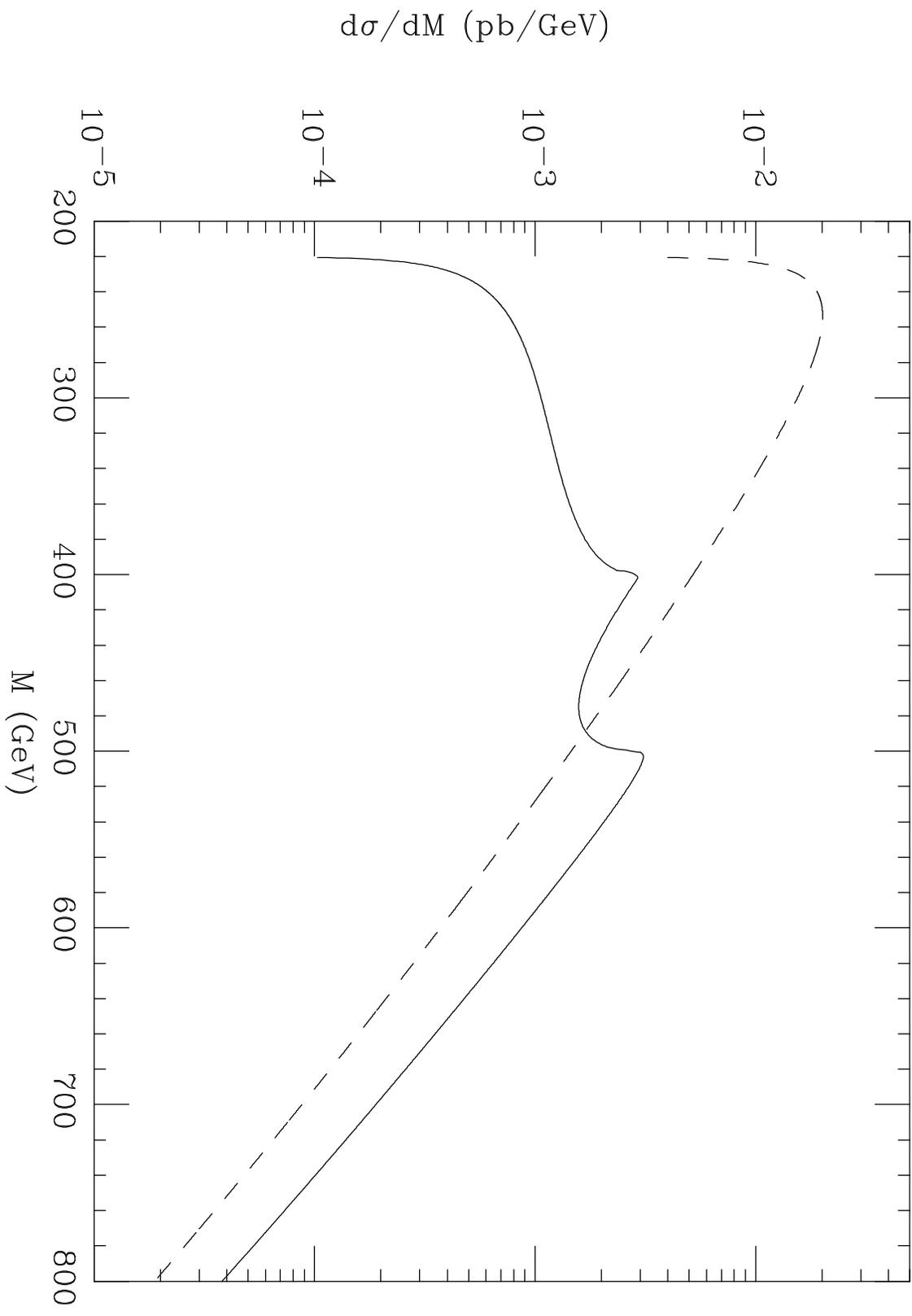
\bye